\pdfoutput=1

\renewcommand{\baselinestretch}{1}

\documentclass[9pt, conference]{IEEEtran}
\usepackage{balance}
\usepackage{nopageno}
\usepackage{float}
\usepackage{tabularx}
\usepackage{mathtools}
\usepackage{amsmath}
\usepackage{hhline}
\usepackage{silence}
\usepackage{graphicx}
\usepackage{soul}
\usepackage{multirow}
\usepackage{graphicx}
\usepackage{multirow}
\usepackage{amsmath}
\usepackage[english]{babel}
\usepackage{booktabs}
\usepackage{array}
\usepackage{paralist}
\usepackage{threeparttable}
\usepackage{lipsum}
\usepackage{flushend}
\usepackage{cuted}
\usepackage{algpseudocode}
\usepackage{algorithm}
\usepackage{amssymb}
\usepackage{multicol}
\usepackage{subfig}
\usepackage{cite}
\usepackage{makecell}
\usepackage{url}
\usepackage[table]{xcolor}
\usepackage[normalem]{ulem}
\newcommand{\ignore}[1]{}
\newcommand{\redHL}[1]{\textcolor{red}{#1}}
\newcommand{\blueHL}[1]{{\textcolor{blue}{#1}}}

\newcommand\blfootnote[1]{%
  \begingroup
  \renewcommand\thefootnote{}\footnote{#1}%
  \addtocounter{footnote}{-1}%
  \endgroup
}

\usepackage{listings}
\begin{document}

\title{Accelerating OTA Circuit Design: Transistor Sizing Based on~a Transformer Model and Precomputed Lookup Tables}

% \author{\IEEEauthorblockN{Subhadip Ghosh}
% \IEEEauthorblockA{\textit{University of Minnesota}\\
% Minneapolis, MN, USA}
% \and
% \IEEEauthorblockN{Endalk Y. Gebru}
% \IEEEauthorblockA{\textit{University of Minnesota}\\
% Minneapolis, MN, USA}
% \and
% \IEEEauthorblockN{Chandramouli Kashyap}
% \IEEEauthorblockA{\textit{Cadence Design Systems}\\
% Portland, OR, USA}
% \and
% \IEEEauthorblockN{Ramesh Harjani}
% \IEEEauthorblockA{\textit{University of Minnesota}\\
% Minneapolis, MN, USA}
% \and
% \IEEEauthorblockN{Sachin S. Sapatnekar}
% \IEEEauthorblockA{\textit{University of Minnesota}\\
% Minneapolis, MN, USA}
% }

\author{
Subhadip Ghosh$^1$, Endalk Y. Gebru$^1$, Chandramouli V. Kashyap$^2$, Ramesh Harjani$^1$, Sachin S. Sapatnekar$^1$ \\
$^1$\text{Department of Electrical and Computer Engineering, University of Minnesota}, Minneapolis, MN, USA \\
$^2$\text{Cadence Design Systems}, Portland, OR, USA
}

\maketitle

\begin{abstract}
Device sizing is crucial for meeting performance specifications in operational transconductance amplifiers (OTAs), and this work proposes an automated sizing framework based on a transformer model. The approach first leverages the driving-point signal flow graph (DP-SFG) to map an OTA circuit and its specifications into transformer-friendly sequential data. A specialized tokenization approach is applied to the sequential data to expedite the training of the transformer on a diverse range of OTA topologies, under multiple specifications. Under specific performance constraints, the trained transformer model is used to accurately predict DP-SFG parameters in the inference phase. The predicted DP-SFG parameters are then translated to transistor sizes using a precomputed look-up table-based approach inspired by the $g_m/I_d$ methodology. In contrast to previous conventional or machine-learning-based methods, the proposed framework achieves significant improvements in both speed and computational efficiency by reducing the need for expensive SPICE simulations within the optimization loop; instead, almost all SPICE simulations are confined to the one-time training phase. The method is validated on a variety of unseen specifications, and the sizing solution demonstrates over 90\% success in meeting specifications with just one SPICE simulation for validation, and 100\% success with 3--5 additional SPICE simulations. 
% \redHL{Can we quantify ``small'' -- e.g., ``after 3-5 iterations''? Can we change ``iteration'' to ``SPICE simulation''?\textbf{YES, How about ``additional'' with that?}}

\end{abstract}

\vspace{-1mm}
\section{Introduction}
\label{sec:Intro}

\noindent

\noindent
A ``perfect storm'' of events is driving the need for greater automation in analog design: increased demand for analog circuits within larger systems; a diminishing designer workforce that is hard-pressed to fulfill this demand; and increased design complexity due to the complexity of circuit models in advanced nodes. Today, it is imperative to develop automation techniques that improve designer productivity.\blfootnote{This work was supported in part by NSF (award 2212345) and SRC.} 

One of the most critical blocks in many analog systems is the operational transconductance amplifier (OTA), which performs tasks such as amplification, filtering, and signal conditioning.  Transistor sizing for OTAs has long been a time-consuming process, relying on human experts to optimize circuit metrics under strict performance constraints. Early automated approaches employed knowledge-based methods~\cite{harjani_89}, but codifying expertise into exhaustive rules is challenging, especially with the need for updates across technology nodes. 

% Equation-based approaches in~\cite{abel_22, abel_22_2} are suitable for older technology nodes, but design in more recent nodes involves more complex device models and SPICE simulation. \redfn{Can remove some if required. I am not sure if I should cite multiple papers from the same author. Can keep 2, remove 1. Use your judgment on which ones to keep. And I still see inconsistencies in the refs (e.g., [7] provides an online link for some reason. The ref list needs more careful inspection.)}, yet they often become intractable for complex topologies\bluefn{Not a good way to criticize this. You have the same failings for the results you show.} and deviate from actual performance. 
% \blueHL{\sout{Methods based on stochastic search, such as genetic algorithms~\cite{Kruiskamp_95}, simulated annealing~\cite{gielen_90}, evolutionary algorithms~\cite{liu_09}, and particle swarm optimization~\cite{vural_12}, rely heavily on SPICE simulations, leading to slow optimization and challenges with convergence.}}
% \bluefn{Again, you have messed up my corrections here. I had SPECIFICALLY pointed out that ``heavily'' is not appropriate usage in formal writing. \textbf{Ok Prof.}}
Methods based on stochastic search (genetic algorithms~\cite{koza_96, Kruiskamp_95}, simulated annealing~\cite{gielen_90}, 
and particle swarm optimization~\cite{vural_12}) have been explored for OTA optimization, but accuracy requires numerous expensive SPICE simulations within the optimization loop.
Equation-based techniques~\cite{abel_22, abel_22_2} and approaches based on convex optimization, such as geometric programming using posynomial-form models~\cite{hershenson_01}, work for simplified MOS models, but can falter in the face of the complex device models in advanced nodes. 
% Methods such as\bluefn{NEVER use ``like'' in formal writing. Use ``such as'' instead. I am pretty sure I have said this before to you. Please search and replace all such instances. If I work at such low-level issues, I will never get around to giving you useful feedback. I am a professor, not a secretary. Use my abilities as a professor, not as a proofreader for your errors and careless work.} GASPAD~\cite{liu_14}, developed for mm wave IC synthesis,\bluefn{This appears to be complete nonsense. Does GASPAD use OTAs???? What are you thinking? And can you not take the trouble to fix the references? Why do you have ``Gaspad'' in the references instead of ``GASPAD''?? I have laboriously fixed all your errors, but you put in zero effort on your part and ruin the reference section with you poor work. This really is very poor.} use surrogate models such as Gaussian process regression (GPR)~\cite{rasmussen_2004}. While GPR has advantages, its cubic computational complexity, $O(N^3)$, limits its scalability as the sample size grows. Additionally, Bayesian optimization has been used with GPR ~\cite{lyu_18}, but they require numerous circuit simulations, rendering them computationally inefficient.\bluefn{I am not sure how much of this paragraph makes sense beyond the description of GASPAD. Let me come back to this later. If you can fix it, it will be helpful. Please, use your intelligence. Why are you talking about mm-wave circuits in a paper about a low-frequency circuit such as an OTA???? If you leave your brains at home, you may as well not bother writing a paper.}
Recent works have proposed ML-based techniques such as DNN-Opt~\cite{budak_21}, AutoCkt~\cite{settaluri_20}, and GCN-RL~\cite{Wang_2020}, and an RL approach using
% $g_m/I_D$ method~\cite{silviera_96} for
sensitivity analysis~\cite{choi_23}. 
% While these methods reduce the number of simulations compared to older black-box approaches, 
For these methods, OTA sizing under each new set of performance specifications  requires numerous SPICE simulations.
% \bluefn{Please check that these all handle OTAs. There is zero effort in the writeup to link any of these references to OTAs. At this point, I give up. I have no time to fix this, and I have no time to fix any ``corrections'' you may make, because those are poorly thought through and require substantial effort on my part. Let's go with this since it seems to be the best you want to do, and you don't think it is worth doing better. I give up. I cannot teach you to WANT to do better. I will just do the secretarial work of fixing your errors. Apparently that is what you need because you don't want to put any effort into doing better.\textbf{No Prof, I will do a quick check.}}
% \bluefn{Unclear. You also face challenges due to circuit topology -- you need to retrain for each topology. So why are these methods being criticized here? \textbf{The biggest drawback of these approaches, which we are solving is SPICE dependency. So, I think I should focus on that only. Even my approach is topology dependent. I commented that part.}} SSS_NOTE 

We present a transistor sizing approach for OTA circuits using a transformer architecture~\cite{vaswani_17}. This method employs an attention mechanism to capture complex nonlinear relationships between device parameters and circuit performance, addressing the intricacies of modern technology nodes. A key feature of our approach, in contrast with prior methods, is that the cost of SPICE simulations is confined to a one-time training phase. \textit{In contrast with prior methods, for a given set of specifications, in the inference phase, the sizing solution can be obtained rapidly with very few SPICE simulations.}

The transformer requires the circuit characteristics to be represented as a character sequence that forms a set of input tokens to the transformer. We achieve this by utilizing the driving-point signal flow graph (DP-SFG) from~\cite{ochoa_98}, later modified in~\cite{schmid_18}, to facilitate direct mapping of the schematic to a graph. We note that the DP-SFG representation of a circuit serves as a descriptive language, translating the behavior of the circuit into a character string that encapsulates its parameters and structure. Based on this observation, we map the transistor sizing problem to a language-translation task akin to natural language processing (NLP). For a given query representing desired performance specifications, the transformer is trained to output a string with the DP-SFG parameter values that meet these specifications. We translate DP-SFG parameters to predict the transistor sizes based on precomputed look-up tables (LUTs). 

An ML-based approach does not guarantee perfect accuracy, and occasionally, the predicted design point may show minor violations in the specifications in some test cases. 
% \redHL{I had specifically changed the language to fix these sentences. Do you have the old version that I had fixed? Please do not overwrite my changes and make me duplicate effort. {\em That is not it. I had removed the part that said ``No ML-based approach...'' Why do you overwrite my changes when you don't know how to write?? I will have to do this again.} \textbf{PROF, I won't do that again.} You are going to cause me to have a late night because of your thoughtlessness. I am not at all happy.\textbf{Prof, this will never happen again.} It's ok to change, but at least don't DELETE my changes. Put them in comments, or save a copy, or do something that allows recoverability. Now I have to waste my time on doing this all over again.\textbf{I will keep that in mind.} 
% \redHL{I will keep this as is. I don't really have time to fix it now, and while the new wording would have made this better, nothing that is stated here is incorrect.} 
In these instances, the designer can re-invoke the fast transformer-based method with tighter design specifications until all requirements are met. Thus, the method acts as a copilot, keeping the designer in the loop. 

The key contributions of our work are as follows:
\vspace{-0.05cm}
\begin{itemize}
\setlength\itemsep{0.05cm}
    \item We implement an automated framework to map an OTA circuit to its equivalent DP-SFG. The paths of this DP-SFG are encoded and concatenated in a language that describes circuit behavior.

    \item We create a labeled training set for our ML model, corresponding to a range of device sizes, over multiple OTA topologies, and evaluate performance metrics using SPICE simulations. 
    
    \item We customize a transformer-based encoder-decoder to encode paths in the DP-SFG
    % ; based on the training data, the transformer 
    to predict the DP-SFG parameters such as transconductance $(g_m)$ and capacitance values.
    
    \item We utilize our LUTs, precharacterized using SPICE simulations, to translate the outputs of the transformer into device sizes, thus providing the sizing solution that meet specifications.
\end{itemize}
Our approach is flexible enough to be used within a layout optimization loop. After sizing, a layout engine updates parasitics, updating the parasitic values in the DP-SFG.  Our model, trained on a range of values, can then be re-invoked without further SPICE simulations.

The paper is structured as follows. Section~\ref{sec:background} overviews a set of core building blocks used in our approach, and is followed by Section~\ref{sec:sizing_framework}, which details our proposed sizing framework and its implementation. Next, Section~\ref{sec:results} details our experimental setup and findings, and finally, Section~\ref{sec:conclusion} concludes the paper.
\section{Background}
\label{sec:background}

\noindent
In this section, we first overview the principles governing transformer architecture. Next, we present a concise overview of DP-SFGs, which we employ to map OTA circuits into transformer-friendly sequential data. Finally, we describe a precomputed LUT-based width estimator to translate DP-SFG parameters to transistor widths.
\vspace{-1mm}
\subsection{The transformer architecture}

\noindent
The transformer~\cite{vaswani_17} is viewed as one of the most promising deep learning architectures for sequential data prediction in NLP.  It relies on an attention mechanism that reveals interdependencies among sequence elements, even in long sequences. The architecture takes a series of inputs \((x_1, x_2, x_3, \ldots, x_n\)) and generates corresponding outputs \((y_1, y_2, y_3, \ldots, y_n\)).

\begin{figure}[b]
\vspace{-5mm}
\centering
\includegraphics[width=0.5\textwidth, bb=0 0 370 190]{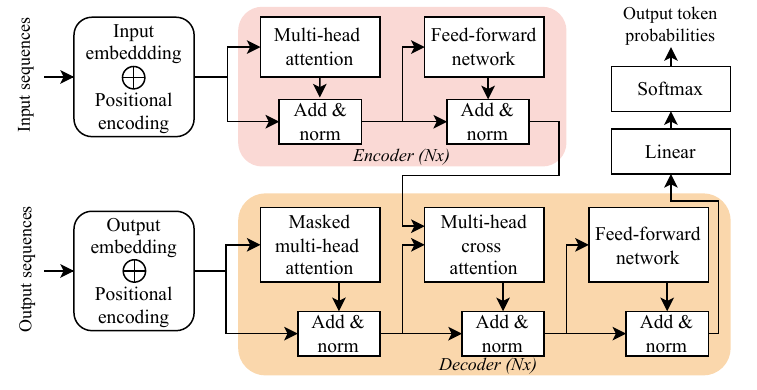}
\vspace{-5mm}
\caption{Architecture of a transformer.}
\label{fig:simpleTrans}
% \vspace{-2mm}
\end{figure}

The simplified architecture shown in Fig.~\ref{fig:simpleTrans} consists of $N$ identical stacked encoder blocks, followed by $N$ identical stacked decoder blocks. The encoder and decoder is fed by an input embedding block, which converts a discrete input sequence to a continuous representation for neural processing. Additionally, a positional encoding block encodes the relative or absolute positional details of each element in the sequence using sine-cosine encoding functions at different frequencies. This allows the model to comprehend the position of each element in the sequence, thus understanding its context. Each encoder block comprises a multi-head self-attention block and a position-wise feed-forward network (FFN); each decoder block, which has a similar structure to the encoder, consists of an additional multi-head cross-attention block, stacked between the multi-head self-attention and feed-forward blocks. The attention block tracks the correlation between elements in the sequence and builds a contextual representation of interdependencies using a scaled dot-product between the query ($Q$), key ($K$), and value ($V$) vectors:
\begin{equation}
\text{{Attention}}(Q, K, V) = \text{softmax}\left(\frac{QK^T}{\sqrt{d_k}}\right)V,
\end{equation}
where $d_k$ is the dimension of the query and key vectors. The FFN consists of two fully connected networks with an activation function and dropout after each network to avoid overfitting. The model features residual connections across the attention blocks and FFN to mitigate vanishing gradients and facilitate information flow.

\subsection{Driving-point signal flow graphs}

\noindent
The input data sequence to the transformer must encode information that relates the parameters of a circuit to its performance metrics.  Our method for representing circuit performance is based on the signal flow graph (SFG).  The classical SFG proposed by Mason~\cite{Mason53} provides a graph representation of linear time-invariant (LTI) systems, and maps on well to the analysis of linear analog circuits such as amplifiers. In our work, we employ the driving-point signal flow graph (DP-SFG)~\cite{ochoa_98,schmid_18}. The vertices of this graph are the set of excitations (voltage and current sources) in the circuit and internal states (e.g., voltages) in the circuit.  
% An edge is drawn between vertices that have an electrical relationship, and the weight on each edge is the gain of the edge;
An edge connects vertices with an electrical relationship, and the edge weight is the gain; 
for example, if a vertex $z$ has two incoming edges from vertices $x$ and $y$, with gains $a$ and $b$, respectively, then $z = ax + by$, using the principle of superposition in LTI systems.  To effectively use superposition to assess the impact of each node on every other node, the DP-SFG introduces auxiliary voltages at internal nodes of the circuit that are not connected to excitations. These auxiliary sources are structured to not to alter any currents or voltages in the original circuit, and simplifies the SFG formulation for circuit analysis.
% enable easy formulation of the SFG to analyze circuit behavior. 

\begin{figure}[t]
% \vspace{-6mm}
\centering
\includegraphics[width=0.9\linewidth, bb=0 0 320 140]{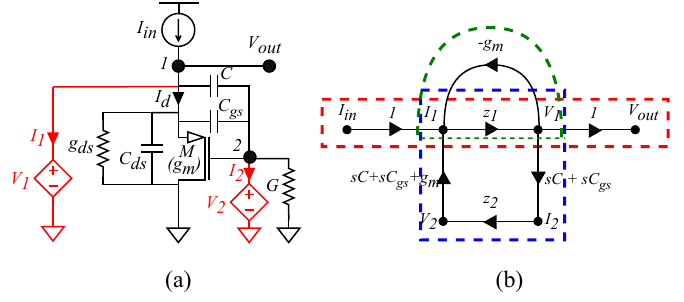}
\vspace{-0.25cm}
\caption{~(a) Schematic and (b) DP-SFG for an active inductor.}
\label{fig:DP-SFG_ex}
\vspace{-5mm}
\end{figure}

Fig.~\ref{fig:DP-SFG_ex}(a) shows a circuit of an active inductor, which is an inductor-less circuit that replicates the behavior of an inductor over a certain range of frequencies. Fig.~\ref{fig:DP-SFG_ex}(b) shows the equivalent DP-SFG. In Section~\ref{sec:dp-sfg}, we provide a detailed explanation that shows how a circuit may be mapped to its equivalent DP-SFG.

\section{Proposed Methodology}
\label{sec:sizing_framework}

\subsection{Overview of the solution}

\noindent
We leverage transformer models to capture the complex relationships between device attributes and circuit performance. We conceptualize the transistor sizing problem as a language translation task, where the input sequence consists of a DP-SFG representation for an OTA circuit, together with performance specifications. The transformer model generates an enhanced DP-SFG representation with the device characteristics necessary to meet the specifications.

\begin{figure}[b]
  \vspace{-5mm}
  \centering
  \includegraphics[width=0.85\linewidth,bb=0 0 295 180]{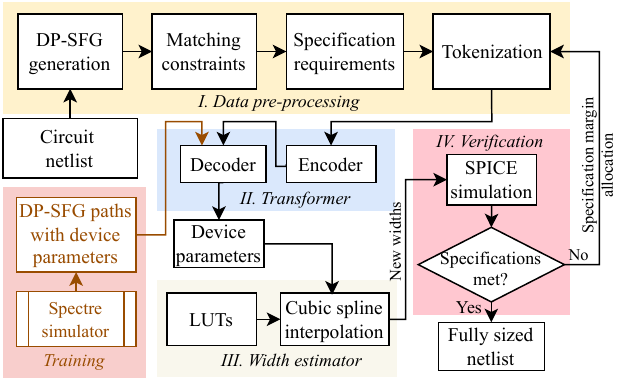} % Replace example-image with your image file
  % \vspace{-0.5cm}
  \caption{Overall sizing flow using our transformer-based method.}
  \label{fig:toplevel}
  % \vspace{-2mm}
\end{figure}

Fig.~\ref{fig:toplevel} illustrates the workflow of our framework, with four stages. Stage I performs preprocessing, generating the DP-SFG from the circuit netlist.  The DP-SFG and the designer-specified performance constraints are then tokenized into a combined sequence. Next, in Stage II, a transformer model processes these tokens to predict circuit parameters that meet performance specifications; these are then translated to individual device widths in Stage III using the precomputed LUTs and a $g_m/I_d$ methodology. Finally, in Stage~IV, the performance of the predicted sized circuit is verified using SPICE simulation. In a vast majority of cases, we will show that the performance criteria are satisfied; if not, the designer is brought into the loop to provide tighter specifications, and procedure is reinvoked so that the original specifications are met. The remainder of this section discusses the detailed implementation of each stage.

\vspace{-2mm}
\subsection{Circuit-to-sequence mapping using the DP-SFG}
\label{sec:dp-sfg}

\noindent
The procedure for creating the DP-SFG formalizes the approach in~\cite{schmid_18,schmid_yt}. We use the running example of an active inductor circuit, whose DP-SFG is shown in Fig.~\ref{fig:DP-SFG_ex}(b) to illustrate the method.

\noindent
\textbf{Step 0: Initial bookkeeping and node initialization}
The algorithm begins with initializing the vertex (or node) set $V$, and initializing data structures for fast access to connectivity information between circuit components (RCs, transistors, etc.) from the netlist. 

\noindent
\textbf{Step 1: Insertion of auxiliary nodes.} 
For nodes that are not connected to voltage sources, we create auxiliary voltage sources. These sources are described by $V = z I$, where $z$ is the driving point impedance (DPI) at the node, i.e., the the inverse sum of all conductances connected to the node. In Fig.~\ref{fig:DP-SFG_ex}, auxiliary sources are added at nodes~1 and~2. The sources replicate node voltages without introducing additional current into the circuit.  This establishes relationships $V_1 = z_1 I_1 $ and $V_2 = z_2 I_2$, where 
\begin{equation}
z_1 = \frac{1}{sC+sC_{\textit{ds}}+sC_{\textit{gs}}+g_{\textit{ds}}}, \; \;
z_2 = \frac{1}{sC+sC_{\textit{gs}}+G}
\end{equation}

\noindent
\textbf{Step 2: Adding branches due to passive components.} Next, we add the edges associated with passive components. We consider how each terminal connects to auxiliary sources. If a terminal connects to an auxiliary source, we connect it to the auxiliary node of the other terminal using its admittance as the edge weight. If neither terminal connects to an auxiliary source, we connect them with an edge using the admittance of the component. In Fig.~\ref{fig:DP-SFG_ex}, the terminals of capacitor \( C \) are both connected to auxiliary sources carrying currents \( I_1=s(C + C_{\textit{gs}}) V_2 \) and \( I_2=s(C + C_{\textit{gs}}) V_1 \) through the associated edges.

\noindent
\textbf{Step 3: Adding branches due to transistor transconductances.} 
The voltage at each terminal of a transistor influences its drain current. Based on these terminal voltages, we establish connections that directly or indirectly affect the auxiliary node currents. In the example of Fig.~\ref{fig:DP-SFG_ex}, if $V_1$ increases, the drain current $I_d$ increases, and hence current flowing to $I_1$ decreases in the opposite direction. This is reflected by setting the weight on the branch from $V_1$ to $I_1$ to $-g_m$. Similarly, the branch $V_2$ to $I_1$ has weight $+g_m$, reflecting the positive dependence of drain current $I_d$ with $V_2$. 

\begin{figure}[t]
  \centering
  \includegraphics[width=0.9\linewidth, bb=0 0 240 170]{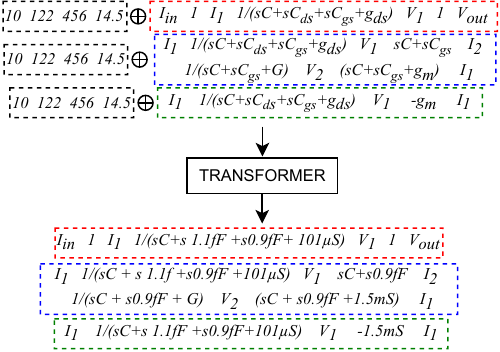} % Replace example-image with your image file
  \caption{\textbf{Input:} DP-SFG paths with desired performance specifications, \textbf{Output:} Predicted sequences with device parameter values.}
  \label{fig:seq}
  \vspace{-0.4cm}
\end{figure}

At the end of Step~3, we obtain the final DP-SFG in Fig.~\ref{fig:DP-SFG_ex}(b) for the active inductor circuit. We will encode such a DP-SFG into sequential data that encapsulates the functionality of the circuit as well as the parameters of circuit components, including parasitics. This sequence representation is provided as an input to the transformer and is eventually used to size transistors in the circuit. We utilize the NetworkX Python package to process the final DP-SFG. This approach utilizes Johnson's algorithm ($O(V^2 \log V + VE)$ complexity) to identify all cycles, and the depth-first search algorithm ($O(V + E)$ complexity) to find all forward paths, where $V$ represents the number of nodes and $E$ represents the number of edges in the graph. 
For our running example, Fig.~\ref{fig:seq} shows the sequences obtained from the DPSFG. Specifically, the path outlined by red dotted rectangle represents the forward path between input and output nodes, while the blue and green outlined paths denote the cycles.

\subsection{Implementation of the transformer}

\noindent
\textbf{Transformer inputs.} The transformer model comprehends the interdependencies between device parameters and circuit performance metrics. We frame the paths by concatenating the nodes and edge weights from the DPSFG for every forward path and loop. The transformer takes in the list of paths extracted from the DP-SFG, each augmented with the desired set of specifications outlined by the black dotted rectangle in Fig.~\ref{fig:seq}. The transformer acts on the sequences and predicts the device parameters $g_m$, $g_{ds}$, $C_{ds}$, and $C_{gs}$ which will satisfy the desired specifications.

\noindent
\textbf{Overall transformer architecture.}
We implement the transformer in Python, leveraging the PyTorch library. The architecture of the transformer model closely resembles the one proposed by~\cite{vaswani_17}, with minor modifications. We use a 720-dimensional input embedding with 12 heads of parallel attention layers, while keeping the rest of the parameters unchanged.

\noindent
\textbf{Tokenization and byte-pair encoding}.
Tokenization is a crucial step for optimizing transformer efficacy. It breaks down a long sequence into smaller entities called tokens. Unlike in traditional NLP, where individual words and sub-words are treated as tokens, we use specific groups of individual characters such as the key device parameters $g_m, g_{ds}$, $C_{ds}$, $C_{gs}$, edge weights, and the names of the transistors, as tokens that convey necessary information about the circuit. 

For our problem, character-level tokenization (CLT), where each character is treated as a single token, is found to lead to long sequence lengths, i.e., a large number of tokens within a single sequence, resulting in computational inefficiency.
To overcome this problem, we employ the byte-pair encoding (BPE) approach~\cite{rico_16}.
This approach iteratively combines the most frequently occurring tokens (bytes) into a single token,
and dynamically adapts the vocabulary of the training data to capturing a common group of characters conveying essential information. By applying BPE, we achieve a 3.77$\times$ compression of the sequence lengths compared to the use of CLT, 
% \bluefn{Compared to what baseline? CLT? This is meaningless without stating the baseline. \textbf{FIXED}} 
leading to substantial savings in training time and memory requirements.

To demonstrate tokenization and for an actual DP-SFG path, we choose a partial path of a five-transistor operational transconductance amplifier (5T-OTA). The results of CLT and BPE are:
% character-level tokenization $(A)$ and BPE $(B)$.

\noindent
% \textit{Character-level encoding:} Character-level coding treats every single character as an independent individual token, as shown below:\\ 
CLT: \textcolor{white}{\sethlcolor{blue}\hl{3}\sethlcolor{red}\hl{2} \sethlcolor{orange}\hl{g}\sethlcolor{teal}\hl{m}\sethlcolor{violet}\hl{P}\sethlcolor{brown}\hl{1} \sethlcolor{red}\hl{-}\sethlcolor{pink}\hl{1}\sethlcolor{gray}\hl{6}
\sethlcolor{blue}\hl{1}\sethlcolor{black}\hl{/}\sethlcolor{cyan}\hl{(}\sethlcolor{orange}\hl{g}\sethlcolor{lightgray}\hl{d}\sethlcolor{blue}\hl{s}\sethlcolor{teal}\hl{M}\sethlcolor{purple}\hl{0}\sethlcolor{pink}\hl{+}\sethlcolor{magenta}\hl{s}\sethlcolor{teal}\hl{C}\sethlcolor{lightgray}\hl{d}\sethlcolor{brown}\hl{s}\sethlcolor{red}\hl{M}\sethlcolor{blue}\hl{0}\sethlcolor{pink}\hl{+}\sethlcolor{gray}\hl{s}\sethlcolor{black}\hl{C}\sethlcolor{lightgray}\hl{d}\sethlcolor{orange}\hl{s}\sethlcolor{violet}\hl{P}\sethlcolor{blue}\hl{1}\sethlcolor{pink}\hl{+}\sethlcolor{orange}\hl{g}\sethlcolor{teal}\hl{m}\sethlcolor{violet}\hl{P}\sethlcolor{blue}\hl{1}\sethlcolor{cyan}\hl{)}} 

% \textcolor{white}{\sethlcolor{blue}\hl{32} \sethlcolor{orange}\hl{gmP1} \sethlcolor{red}\hl{-16}
% \sethlcolor{blue}\hl{1/(}\sethlcolor{orange}\hl{gdsM0}\sethlcolor{pink}\hl{+}\sethlcolor{lightgray}\hl{s}\sethlcolor{teal}\hl{CdsM0}\sethlcolor{pink}\hl{+}\sethlcolor{lightgray}\hl{s}\sethlcolor{black}\hl{CdsP1}\sethlcolor{pink}\hl{+}\sethlcolor{orange}\hl{gmP1}\sethlcolor{cyan}\hl{)}} 

\noindent
BPE: 
% \textcolor{white}{\sethlcolor{blue}\hl{3}\sethlcolor{pink}\hl{2}\sethlcolor{pink} \hl{2}\sethlcolor{orange}\hl{.}\sethlcolor{violet}\hl{5}\sethlcolor{black}\hl{m}\sethlcolor{magenta}\hl{S}\sethlcolor{orange}\hl{P}\sethlcolor{cyan}\hl{1} \sethlcolor{red}\hl{-}\sethlcolor{cyan}\hl{1}\sethlcolor{brown}\hl{6}
% \sethlcolor{cyan}\hl{1}\sethlcolor{orange}\hl{/}\sethlcolor{red}\hl{(}\sethlcolor{violet}\hl{5}\sethlcolor{black}\hl{6}\sethlcolor{cyan}\hl{7}\sethlcolor{teal}\hl{u}\sethlcolor{magenta}\hl{S}\sethlcolor{violet}\hl{M}\sethlcolor{olive}\hl{0}\sethlcolor{pink}\hl{+}\sethlcolor{magenta}\hl{s}\sethlcolor{olive}\hl{0}\sethlcolor{orange}\hl{.}\sethlcolor{cyan}\hl{7}\sethlcolor{black}\hl{a}\sethlcolor{red}\hl{F}\sethlcolor{teal}\hl{M}\sethlcolor{olive}\hl{0}\sethlcolor{pink}\hl{+}\sethlcolor{magenta}\hl{s}\sethlcolor{violet}\hl{5}\sethlcolor{red}\hl{4}\sethlcolor{gray}\hl{1}\sethlcolor{black}\hl{a}\sethlcolor{red}\hl{F}\sethlcolor{darkgray}\hl{P}\sethlcolor{cyan}\hl{1}\sethlcolor{pink}\hl{+}\sethlcolor{pink}\hl{2}\sethlcolor{orange}\hl{.}\sethlcolor{violet}\hl{5}\sethlcolor{black}\hl{m}\sethlcolor{magenta}\hl{S}\sethlcolor{orange}\hl{P}\sethlcolor{cyan}\hl{1}\sethlcolor{purple}} 
\textcolor{white}{\sethlcolor{blue}\hl{32}\sethlcolor{pink} \hl{2}\sethlcolor{orange}\hl{.}\sethlcolor{violet}\hl{5}\sethlcolor{black}\hl{mS}\sethlcolor{orange}\hl{P1} \sethlcolor{red}\hl{-16}
\sethlcolor{blue}\hl{1/(}\sethlcolor{violet}\hl{5}\sethlcolor{black}\hl{6}\sethlcolor{cyan}\hl{7}\sethlcolor{orange}\hl{uS}\sethlcolor{violet}\hl{M0}\sethlcolor{pink}\hl{+}\sethlcolor{magenta}\hl{s}\sethlcolor{pink}\hl{0}\sethlcolor{violet}\hl{.}\sethlcolor{cyan}\hl{7}\sethlcolor{black}\hl{aF}\sethlcolor{teal}\hl{M0}\sethlcolor{pink}\hl{+}\sethlcolor{gray}\hl{s}\sethlcolor{violet}\hl{5}\sethlcolor{red}\hl{4}\sethlcolor{gray}\hl{1}\sethlcolor{black}\hl{aF}\sethlcolor{darkgray}\hl{P1}\sethlcolor{pink}\hl{+}\sethlcolor{pink}\hl{2}\sethlcolor{orange}\hl{.}\sethlcolor{violet}\hl{5}\sethlcolor{black}\hl{mS}\sethlcolor{orange}\hl{P1}} 

% \hfill--$(B)$

\noindent
The CLT sequence colors each neighboring character differently, denoting the tokenization of each unique character. However, CLT cannot easily comprehend the relation to device parameters or device names. The BPE approach overcomes this by iteratively combining frequently-occurring tokens. For example, BPE combines tokens such as
gmP1, gdsM0, and CdsM0 to represent $g_m$ of transistor P1, and $g_{ds}$ and $C_{ds}$ of transistor M0, respectively. Similarly, character-level tokens for the units of circuit parameters (e.g., ``mS'' or ``aF'') are combined into tokens of multiple characters. However, all purely numeric strings are left uncombined, as shown in the BPE sequence. For instance, for the value 2.5mS, which corresponds to the $g_m$ of transistor P1, the tokens representing 2.5 are maintained as character-level tokens, enabling the transformer to predict each digit relative to performance metrics independently, but the two character-level tokens for ``mS'' are combined into a single token. This restricted BPE representation thus enables the transformer model to better comprehend circuit relationships, as compared with CLT.

\noindent
\textbf{Loss Function.}
To enhance the learning of device parameter prediction from specified inputs, we utilize a weighted cross-entropy loss function for the transformer. Each token is treated as a separate class, with the loss function assigning greater importance to classes critical for accurate predictions. We focus on tokens representing numerical values of device characteristics (e.g., $g_m$, $g_{ds}$, $C_{ds}$, and $C_{gs}$), ensuring they receive more attention during training. This approach allows the transformer to grasp the significance of these characteristics and their impact on performance. Our experiments compared unweighted and weighted loss functions with varying weights, revealing that applying a 20\% increased weight on the numerical tokens yielded optimal performance.

\vspace{-0.02cm}
\subsection{Translating circuit parameters to device widths}
\label{sec:precomputedLUTs}

\noindent
After the trained transformer predicts the values of circuit parameters, they must be transformed to device widths. 
In this section, we describe a methodology 
% that uses the LUT from Section~\ref{sec:LUT} 
for this purpose.

\subsubsection{\textbf{Device characterization}}

In older technologies, the square-law model for MOS transistors could be used to perform a translation between circuit parameters and transistor widths, but square-law behavior is inadequate for capturing the complexities of modern MOS transistor models. In this work, we use a precomputed lookup table (LUT) that rapidly performs the mapping to device sizes while incorporating the complexities of advanced MOS models.

\begin{figure}[t]
% \vspace{-0.4cm}
\centering
\includegraphics[height=3cm, , bb=0 0 210 100]{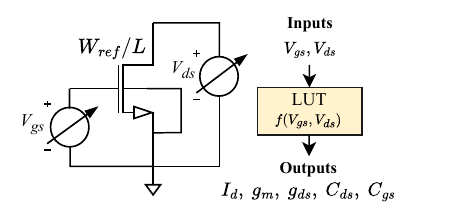}
% \vspace{-0.55cm}
\caption{LUT generation and characterization.}
\label{fig:lutgen}
\vspace{-5mm}
\end{figure}

The LUT is indexed by the $V_{gs}$, $V_{ds}$, and length $L$ of the transistor, and provides five outputs: the drain current ($I_d$), transconductance ($g_m$), drain-source conductance ($g_{ds}$), drain-source capacitance ($C_{ds}$), and gate-source capacitance ($C_{gs}$) all computed per unit transistor width. The entries of the LUT are computed by performing a nested DC sweep simulation across the input indices for the MOSFET with a specific reference width, $W_{ref}$, as shown in Fig.~\ref{fig:lutgen}, and for each input combination, the five outputs are recorded. Since the five quantities all vary linearly with the width of the transistor, we store their corresponding values per unit width. 
% Empirically, we see that the impact of $V_{sb}$ is small enough
% \bluefn{For my info: over what range of $V_{sb}$ have you done this, and why is that sufficient? \textbf{I chose the range between (0 - $V_{dd}/3$), this is due to having around three mosfets b/n $V_{dd}$ and ground in our circuits. For example, in the 5T OTA the $V_{sb}$ for the DP is non-zero since the source is sitting on top of the tail mosfet, and the voltage (Vds) across the tail mos is usually $<$vdd/3}} SSS_NOTE
% that it can be neglected, and therefore we set $V_{sb} = 0$ in the sweeps used to create the LUT. 
The LUT stores the vector-valued function
% \vspace{-2mm}
\begin{align}
[I_d \;\; g_m \;\; g_{ds} \;\; C_{ds} \;\; C_{gs}] = f(V_{gs}, V_{ds})
\end{align}

We have constructed a lookup table for a 65nm technology with a reference transistor width of 700nm, with $V_{gs}$ and $V_{ds}$ values ranging from 0--1.2V with a 60mV step. Given the relatively coarse granularity of data points in the LUT, we have implemented cubic spline interpolation to enhance accuracy at intermediate values. These LUT granularity, together with interpolation, ensures that it provides accurate predictions, and yet has a reasonable size.  

Our methodology uses this LUT, together with the $g_m/I_d$ methodology~\cite{silviera_96,jespers_17}, to translate circuit parameters predicted by the transformer to transistor widths. The cornerstone of this methodology relies on the inherent width independence of the ratio $g_m/I_d$ to estimate the unknown device width: this makes it feasible to use an LUT characterized for a reference width $W_{ref}$.

\subsubsection{\textbf{Width estimation}} 

The width estimation process uses the recorded LUT and transformer-predicted MOSFET parameters to compute the optimal width. The pseudocode for the algorithm employed is presented in Algorithm~\ref{algo:width_estimation}. 
After initialization on line~\ref{algo2:init}, the input is converted to the desired $g_m/I_d$ ratio.  Lines~\ref{algo2:while_begin}--\ref{algo2:while_end} iterate over the LUT to find the $W$ that matches the transformer-supplied parameters. Specifically, line~\ref{algo2:gmId} finds the value of $V_{gs}$ at which the $g_m/I_d$ ratio is met. For this value, lines~\ref{algo2:wcalc_begin}--\ref{algo2:wcalc_end} determine candidate values of $W$, $w_1, \cdots, w_5$, by ratioing $I_d^{in}$, $g_m^p$, $g_{ds}^p$, $C_{ds}^p$, and $C_{gs}^p$, respectively, with the corresponding LUT outputs. We iterate over $V_{ds}$ until $w_1, \cdots, w_5$ are as close as possible. Line~\ref{algo2:while_end} takes a step in this direction using the empirically chosen factor $\alpha = 10^{-4}$. The iterations continue until the candidate width values converge.

    \begin{algorithm}[t]
    \footnotesize
        \caption{Width Estimation}
        \label{algo:width_estimation}
    \begin{algorithmic}[1]
    \State \textbf{Inputs:} Transformer-predicted $g_{m}^p$, $g_{ds}^p$, $C_{ds}^p$, $C_{gs}^p$ and current $I_d^{in}$ for a MOSFET; LUT for the device type (PMOS/NMOS); 
    % reference width, $W_{ref}$, used for recording the LUT; 
    tolerances $\alpha$ and $\epsilon$ 
    \State \textbf{Outputs:} Optimal width, $W$, of the MOSFET
    % DC operating points and width of the MOSFET, namely $V_{gs}^*$, $V_{ds}^*$ and $W^*$
    \State $V_{ds,curr} \gets V_{dd}/2$ , $cost_{curr} \gets \infty$, $\Delta \gets \infty$  \label{algo2:init}
    
    \State $g_m\_I_d  \gets g_{m}^p/I_d$ \textit{// Compute the $g_m\_I_d$ operating point}
    
    \While{$|\Delta| > \epsilon$} \textit{// until current cost $\approx$ previous cost}
        \State mincost$_{prev} \gets$ mincost$_{curr}$, $V_{ds,prev} \gets V_{ds,curr}$ \label{algo2:while_begin}
        % \Statex \hspace{1.3em} \textit{// Getting $V_{gs}$ from the LUT}
        \State Find LUT entry $[I_d \;\; g_m \;\; g_{ds} \;\; C_{ds} \;\; C_{gs}] = f(V_{gs}, V_{ds})$ \label{algo2:gmId}
        \Statex \hspace*{10mm} at which $g_m/I_d = g_m\_I_d$. Report $V_{gs}^p$ for this entry.

        \State At $V_{gs}^p$, the LUT for $[I_d \;\; g_m \;\; g_{ds} \;\; C_{ds} \;\; C_{gs}]$ is $f(V_{ds})$.
        \State \textit{// Find $w_1 \cdots w_5$ as functions of $V_{ds}$}
        \State $w_1$$\gets$$g_{m}^p/g_m$, $w_2$$\gets$$g_{ds}^p/g_{ds}$, $w_3$$\gets$$C_{ds}^p/c_{ds}$, 
        \Statex \hspace*{10mm} $w_4$$\gets$$C_{gs}^p/c_{gs}$, $w_5$$\gets$$I_d^{in}/I_d$

        \Statex \hspace{1.3em} \textit{// Find $V_{ds}$ at which $w_i$s are closest}
        \State cost$(V_{ds}) \gets \sum_{n=1}^{3} \sum_{m=n+1}^{4} |w_n - w_m|$   \label{algo2:wcalc_begin}
        \State mincost$_{curr} = \min_{V_{ds}}$ (cost)
        \State $\Delta \gets$ mincost$_{prev}$ $-$ mincost$_{curr}$ \label{algo2:wcalc_end}
        \Statex \hspace{1.3em} \textit{// Updating the initial guess $V_{ds}$ value}
        \State $V_{ds,curr} \gets V_{ds,curr} + \mbox{sgn}(\Delta).\alpha.V_{ds, prev}$ \label{algo2:while_end}
        
        % \State $g_m = f(V_{gs}) \gets \text{lookup}\_g_{m}(V_{ds,curr},L) $
        % \State $I_d = f(V_{gs}) \gets \text{lookup}\_I_{d}(V_{ds,curr},L) $
        % \State $V_{gs} \gets$ \text{get\ $V_{gs}$ where $g_{m}/I_{d}$ == $g_m\_I_d$} \redHL{Find Vgs corresponding to gm\_Id; lines 9 and 10 provide the function, line 5 provides the value to be looked up}
        % \Statex \hspace{1.3em} \textit{// Reading the four parameters as a function of $V_{ds}$}
        % \State $I_d,g_m,g_{ds},C_{ds} =\ f(V_{ds})\ \gets\ \text{lookup}(V_{gs},L)$   
        
        % \Statex \hspace{1.3em} \textit{// Normalizing the parameters with $W_{ref}$}
        % \State $I_d\_w,\ g_m\_w,\ g_{ds}\_w, c_{ds}\_w\gets f(V_{ds})/W_{ref}$
        % \Statex \hspace{1.3em} \textit{// Width calculation using the normalized functions}
        % \Statex \hspace{1.3em} \textit{// Each width is a function of $V_{ds}$}
        % \State $w_1 \gets g_{m}^p/g_m\_w$
        % \State $w_2 \gets g_{ds}^p/g_{ds}\_w$
        % \State $w_3 \gets C_{ds}^p/c_{ds}\_w$
        % \State $w_4 \gets I_d/I_d\_w$

        % \Statex \hspace{1.3em} \textit{// Determining cost function from the widths}
        % \State $cost\_func \redHL{(Vds)} \gets \sum_{n=1}^{3} \sum_{m=n+1}^{4} |w_n - w_m|$   
        % \State $min\_cost_{curr} = min_{\redHL{all Vds}} (cost\_func)$
        % \State $\Delta \gets min\_cost_{prev}-min\_cost_{curr}$
        % \Statex \hspace{1.3em} \textit{// Updating the initial guess $V_{ds}$ value}
        % \State $V_{ds,curr} \gets V_{ds,curr} + sgn(\Delta).\alpha.V_{ds, prev}$
    \EndWhile
    
    % \State $V_{ds}^* \gets \arg\min(cost\_func)$
    \State $W \gets\ w_1(V_{ds})$
    % \State $V_{gs}^* \gets V_{gs}$    

    \end{algorithmic}
    \end{algorithm}

\subsection{SPICE verification and margin allocation}

\noindent
Finally, we perform just one SPICE simulation to verify compliance with all specifications. If any specification deviates from the requirements, the model modifies the specifications and repeats the inference step to obtain a new set of device sizes. For example, if the gain of the sized OTA is 10\% below the desired value, the model iteratively tightens the specifications to accommodate this 10\% difference in the gain requirement until all specifications are satisfied.
% \bluefn{Do you also do this if the performance is much better than the specification? You would reduce the power/area by using smaller device sizes.  Also: why don't you report power/area anywhere? \textbf{No, I don't do this if the performance is much better than the specifications. However, doing this is very much possible. But, its not guaranteed that reducing size will bring down the performance closer to the specifications.}}SSS_NOTE
\section{Experimental Setup and Results}
\label{sec:results}
\vspace{-2mm}
\noindent
% \blueHL{\sout{In this section, we delve into our experimental setup, covering the process of data generation and pre-processing. Additionally, we present details on model training and validation procedures. Finally, we present the efficacy of our sizing assistant in predicting device sizes with some unseen circuit performance specifications.}}
% \bluefn{Can delete this without much loss of continuity, if we need to save space. \textbf{OK}} SSS_NOTE

\subsection{Data generation and preprocessing}

\begin{figure}[t]
    \centering
    \subfloat[]{
        \includegraphics[width=19.5mm, bb = 0 0 100 110]{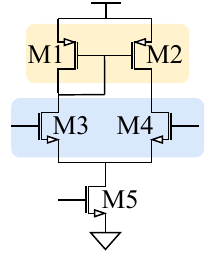}
    }
    \hspace{-2mm} % Adjust spacing between subfigures
    \subfloat[]{
        \includegraphics[width=30.5mm, bb = 0 0 200 110]{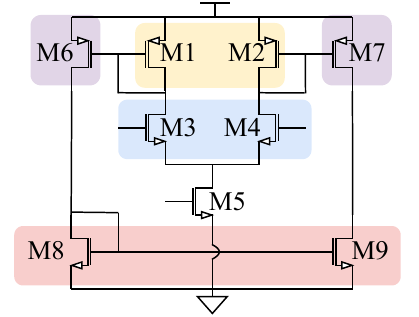}
    }
    \hspace{-1mm} % Adjust spacing between subfigures
    \subfloat[]{
        \includegraphics[width=29mm, bb = 0 0 160 110]{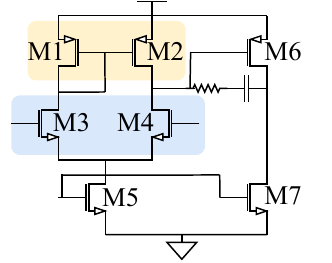}
    }
    
    \caption{Schematic of (a) 5T-OTA, (b) CM-OTA, and (c) 2S-OTA.}
    \label{fig:schemas}
    \vspace{-0.2cm}
\end{figure}

\noindent
To demonstrate the efficacy of our framework, we employ three distinct OTA topologies: five-transistor OTA (5T-OTA), current-mirror OTA (CM-OTA), and two-stage OTA (2S-OTA), each implemented using a 65nm technology node. In Fig.~\ref{fig:schemas}, we show the OTA schematics along with matching constraints under consideration. For clarity in our demonstration, we focus on three performance metrics: gain, 3dB-bandwidth (BW), and unity-gain frequency (UGF), and aim to meet the given performance specifications.
 
Table~\ref{tab:dataset} shows the range of different specifications for the OTAs considered for our training set. We assume that the length of all the devices in a circuit is set to 180nm with a load capacitor $C_L$ of 500fF for all the topologies. To ensure reliable model analysis, we start with precise data generation for each OTA topology using OCEAN scripting. This involves the following steps:
\begin{itemize}
    \item Generating multiple designs with varying transistor sizes by nested sweeps of widths ranging from 0.7$\mu$m to 50$\mu$m.
    \item Enforcing matching constraints for active load current mirror (CM), and differential pair (DP).
    \item Sweeping the DC voltage to determine the input common-mode range (ICMR) of the designs.
    \item Ensuring that the CMs operate in the strong inversion region while the DPs function in the weak inversion region.
    % \bluefn{(1)~Strange choice of words -- what is ``irrelevant''? It either meets specs or it does not. How can it be irrelevant? \textbf{Can I use ``invalid'' ? Cause, I am using ICMR just for filtering out the OTAs which can practically function for a valid input common mode voltage range.} (2)~If you use ICMR as a spec, why isn't it ever reported in your results? \textbf{I am not using ICMR as a SPEC}} SSS_NOTE
    \item Filtering out designs that falls out of the predefined specification range for the dataset outlined in Table~\ref{tab:dataset}.
    \item Capturing the device parameters -- specifically, $g_m, g_{ds}$, $C_{ds}$, and $C_{gs}$ -- for the final legal designs.
\end{itemize}

% \bluefn{What does ``inadequate'' mean? Elaborate. \textbf{I think ``functional'' would be better. The circuits which have a valid ICMR,  and giving out practically useful gain, BW, UGF, are choosen}} SSS_NOTE

Next, we focus on generating appropriate DP-SFG paths for each circuit topology. Table~\ref{tab:dataset} shows the number of sequential paths for each topology. The DP-SFGs are small and the cost of path enumeration is small; for more complex examples, if the number of paths grows large, it is possible to devise other string representations of the DP-SFG.
Finally, in the preprocessing stage, we generate two sets of sequential data, one each for the encoder and the decoder.
% \bluefn{\textbf{***I think at this stage SEQUENTIAL DATA makes more sense***} Is the use of ``sequential paths'' confusing? You are talking about a sequence of tokens that feeds the transformer. The paths are not tokens. Maybe use some other term instead of ``sequential paths''?}
\begin{itemize}
    \item The sequential data at the encoder comprises DP-SFG paths that maintain consistency across all designs within a specific topology. It also includes performance metrics for each design, encompassing gain, BW, and UGF parameters, associated with each unique set of transistor sizes.
    \item The sequential data at the decoder covers the same DP-SFG paths, but with device parameters replaced by values obtained during data generation. These values are unique to each design, aligning with the performance metrics in the encoder sequence.
\end{itemize}

\begin{table}[t]
    \caption{Dataset information.}
    \centering
    % \begin{tabular}{|>{\centering\arraybackslash}m{1.1cm}|>{\centering\arraybackslash}m{1cm}|>{\centering\arraybackslash}m{1.3cm}|>{\centering\arraybackslash}m{1.2cm}|>{\centering\arraybackslash}m{1cm}|>{\centering\arraybackslash}m{0.7cm}|}
    \resizebox{1\linewidth}{!}{\begin{tabular}{|l|c|c|c|c|c|}
        \hline
        \textbf{Topology} & \makecell{\textbf{Gain}\\\textbf{(dB)} \\\textit{min-max}} & \makecell{\textbf{3dB bandwidth}\\\textbf{(MHz)} \\\textit{min-max}} & \makecell{\textbf{UGF} \\\textbf{(MHz)} \\\textit{min-max}}& \makecell{\textbf{\#forward} \\\textbf{paths}} & \textbf{\#cycles} \\
        \hline
        5T-OTA & 18 -- 23 &  7 -- 54 & 80 -- 871  &  9 & 4  \\
        \hline
        CM-OTA & 19 -- 25 & 17.5 -- 86 & 57 -- 1185 & 26 & 5  \\
        \hline
        2S-OTA & 28 -- 54 & 0.01 -- 0.32 & 1.8 -- 370  &  2 & 11 \\
        \hline
    \end{tabular}}
    \label{tab:dataset}
    \vspace{-5mm}
\end{table}

We train a single transformer model that works across all three OTA topologies. By considering all performance criteria and all DP-SFG paths, we convey complete information about each circuit to the transformer. Our dataset comprises 17,000 designs for 5T-OTA, 25,000 designs for CM-OTA, and 8,000 designs for 2S-OTA, each with a different set of transistor sizes. This diverse dataset trains the model across multiple design specification requirements.

% \begin{table}[t]
%     \vspace{-0.2cm}
%     \caption{Dataset Information}
%     \centering
%     \begin{tabular}{|>{\centering\arraybackslash}p{1.1cm}|>{\centering\arraybackslash}p{0.9cm}|>{\centering\arraybackslash}p{0.7cm}|>{\centering\arraybackslash}p{0.5cm}|>{\centering\arraybackslash}p{0.5cm}|>{\centering\arraybackslash}p{0.5cm}|>{\centering\arraybackslash}p{0.5cm}|>{\centering\arraybackslash}p{0.5cm}|>{\centering\arraybackslash}p{0.5cm}|}
%         \hline
%         \multirow{2}{*}{\textbf{Topology}} & \multirow{2}{*}{\textbf{Forward}} & \multirow{2}{*}{\textbf{Cycles}} & \multicolumn{2}{c|}{\makecell{\textbf{Gain}\\\textbf{(dB)}}} & \multicolumn{2}{c|}{\makecell{\textbf{Bandwidth}\\\textbf{(MHz)}}} & \multicolumn{2}{c|}{\makecell{\textbf{UGF} \\\textbf{(MHz)}}}\\
%         \cline{4-9}
%          & \textbf{paths} &  & \textbf{Min} & \textbf{Max} & \textbf{Min} & \textbf{Max} & \textbf{Min} & \textbf{Max} \\
%         \hline
%         5T-OTA & 9 & 4 & 18 & 29 & 7 & 54 & 80 & 871\\
%         \hline
%         CM-OTA & 26 & 5 & 20 & 32 & 59 & 86 & 57 & 1185\\
%         \hline
%         2S-OTA & 2 & 11 & 30 & 52 & 13 & 32 & 18 & 370\\
%         \hline
%     \end{tabular}
%     \label{tab:DP-SFG}
% \end{table}

\subsection{Training and validation}

% \redfn{CK: The number of paths in a more complex circuit could be very large? How do we propose to handle it? \textbf{Since, the inferencing is quite fast, the overall process will still be fast. But, yes the size of the dataset is proportional to number of paths in DPSFG. {\em Added text below Table 1. Please check blueHL.}YES MAKES SENSE}}

\noindent
For our experiments, we employ an Nvidia L40S GPU equipped with 45GB of memory. The dataset is split into an 80:20 ratio for training and validation across each OTA topology. We train a single model using datasets from all three topologies for 40 epochs, employing an adaptive learning rate strategy with the Adam optimizer, beginning with an initial learning rate of $10^{-4}$. Subsequently, our framework is validated against unseen performance specifications across all three OTA topologies. For a given topology and performance specifications, the validation phase rapidly predicts a sequence of tokens corresponding to circuit parameters.The transformer takes in the encoder sequences list and predicts output sequences containing the device parameter values that satisfy the specifications. This is followed by the LUT-based estimator that translates the predicted device parameters to transistor widths. 

% In the subsequent subsection, we comprehensively analyze each aspect of our framework's performance.

\begin{figure}[t]
\vspace{-4mm}
\centering
\subfloat[]{
  \includegraphics[width=0.49\linewidth, bb = 0 0 1100 1000]{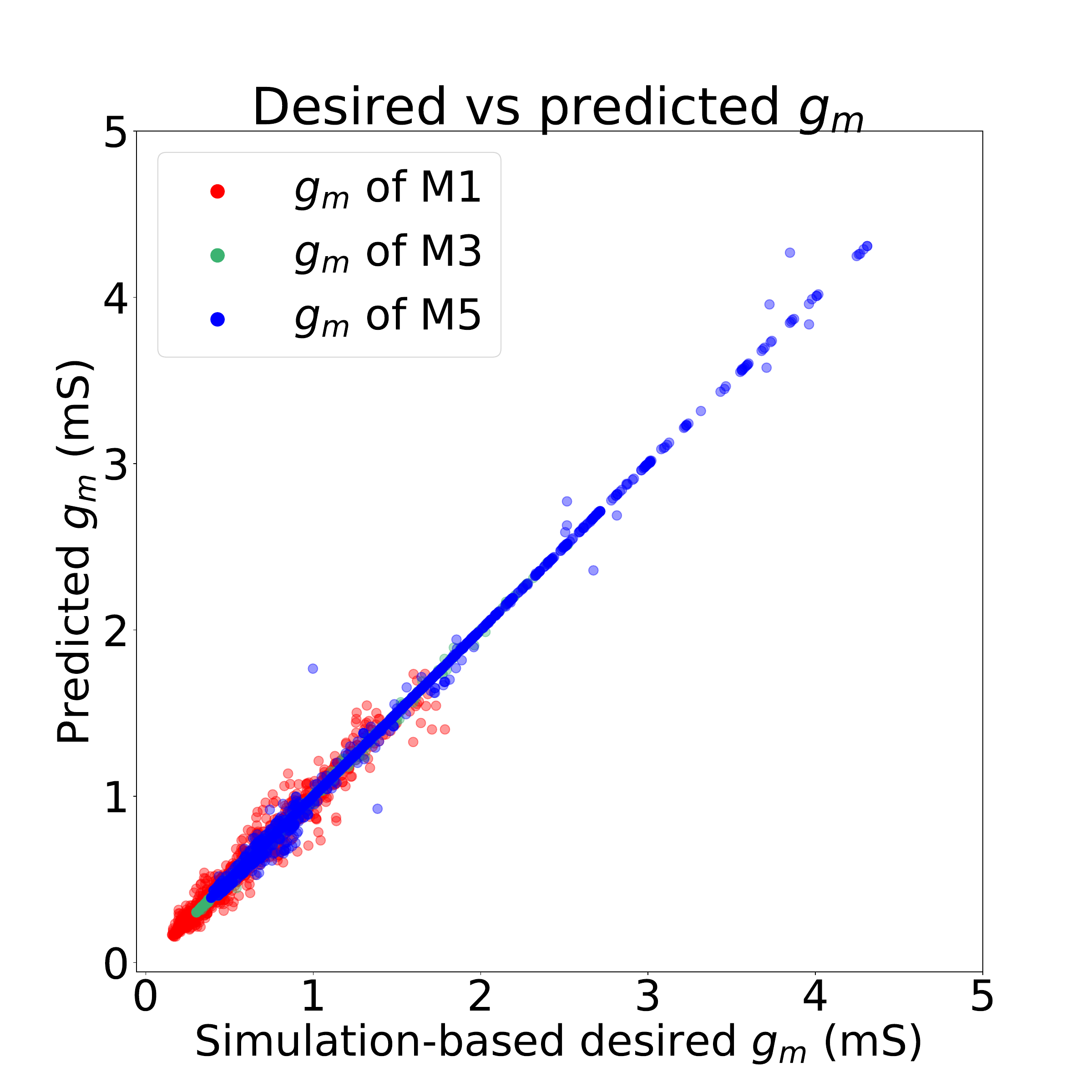}
}
\hspace{-0.4cm}
\subfloat[]{
  \includegraphics[width=0.49\linewidth, bb = 0 0 1100 1000]{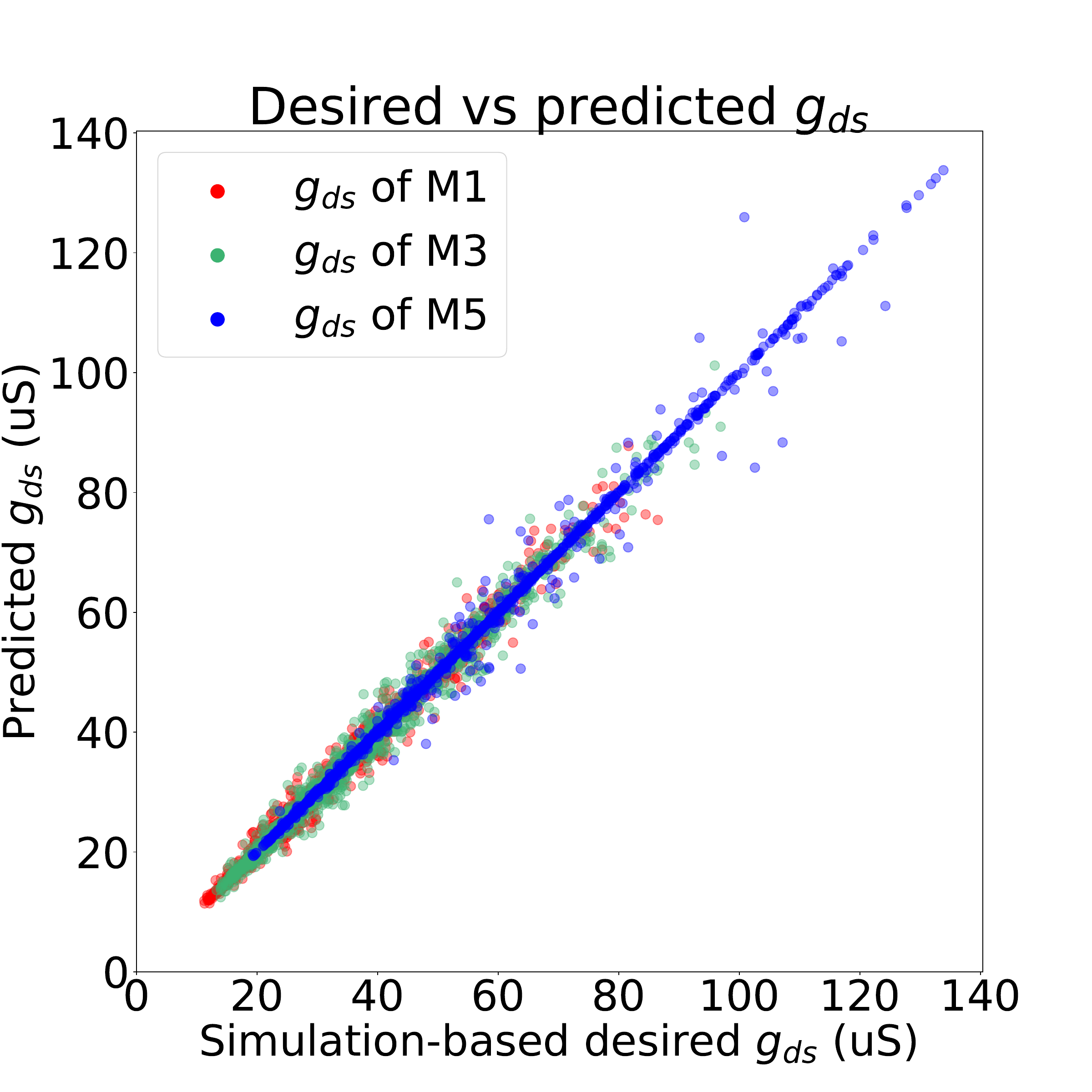}
}
% \vspace{-2mm}
\caption{For 5T-OTA: Scatter plots showing comparisons between predicted and simulation-based device parameters (a) $g_m$ and (b) $g_{ds}$.}
\label{fig:plots}
\vspace{-5mm}
\end{figure}

\subsection{Performance of the framework}

\noindent
We conduct comprehensive performance evaluations to assess the effectiveness of the transformer model and LUT-based width estimator for each OTA topology. Our method sizes 100 unique designs per topology, each with distinct performance specifications not included in the training set. For each specification, the transformer model predicts the key device parameters, which are then converted to transistor sizes using the LUT-based method. The performance of the final design is validated through Spectre simulation of the sized OTA circuit for each topology. Additionally, we ensure the optimized devices operate in the desired region of operation.

\begin{table}[b]
    \vspace{-4mm}
    \centering
    % First table
    \begin{minipage}{\linewidth}
        \centering
        \caption{\centering Correlation coefficient of device parameters between validation data and model outputs for the 5T-OTA.}
        \resizebox{1\linewidth}{!}{
            \begin{tabular}{|>{\centering\arraybackslash}p{0.9cm}|>{\centering\arraybackslash}p{2.3cm}|c|c|c|c|}
                \hline
                \multirow{2}{*}{\makecell{\textbf{MOS} \\ \textbf{devices}}} & \multirow{2}{*}{\makecell{\textbf{Transistor} \\ \textbf{information}}} & \multicolumn{4}{c|}{\makecell{\textbf{Correlation coefficient}}} \\ \cline{3-6}
                & & \textbf{$g_m$} & \textbf{$g_{ds}$} & \textbf{$C_{ds}$} & \textbf{$C_{gs}$} \\
                \hline
                M1/M2 & Active load & 0.982 & 0.993 & 0.962 & 0.964 \\ \cline{1-6} 
                M3/M4 & DP & 0.999 & 0.991 & 0.997 & 0.998 \\ \cline{1-6} 
                M5 & Tail MOS & 0.999 & 0.997 & 0.997 & 0.997 \\ \cline{1-6} 
                \hline
            \end{tabular}
        }
        \label{tab:5t_corr}
    \end{minipage}
    
    \vspace{1mm} % Space between tables

    % Second table
    \begin{minipage}{\linewidth}
        \centering
        \caption{\centering Comparison of optimized design performance with target specifications for the 5T-OTA}
        \resizebox{1\linewidth}{!}{
            \begin{tabular}{|c|c|c|c|c|c|}
                \hline
                \multicolumn{2}{|c|}{\textbf{Gain (dB)}} & \multicolumn{2}{c|}{\makecell{\textbf{UGF (MHz)}}} & \multicolumn{2}{c|}{\makecell{\textbf{3dB bandwidth (MHz)}}} \\ \cline{1-6}
                \makecell{\textbf{Target}} & \makecell{\textbf{Optimized}} & \makecell{\textbf{Target}} & \makecell{\textbf{Optimized}} & \makecell{\textbf{Target}} & \makecell{\textbf{Optimized}} \\
                \hline
                20.13 & 20.6 & 118.78 & 144.64 & 11.38 & 13.33 \\ \cline{1-6} 
                21.23 & 21.37 & 181.25 & 185.38 & 15.31 & 15.49 \\ \cline{1-6} 
                22.78 & 22.79 & 281.75 & 288.54 & 20.18 & 20.48 \\ \cline{1-6} 
                \hline
            \end{tabular}
        }
        \label{tab:5t_specs}
    \end{minipage}

    % \vspace{-2mm}
\end{table}

\noindent
\textbf{5T-OTA}
The 5T-OTA topology includes a matched active current-mirror load (M1/M2), a differential pair (M3/M4), and a tail transistor (M5), all requiring precise sizing to meet performance targets. We assess the prediction accuracy of the transformer by correlating predicted device parameters with SPICE-based validation results. Fig.~\ref{fig:plots} shows a strong correlation between predicted \(g_{m}\) and \(g_{ds}\) values and their SPICE counterparts along the 45° line,
% , where we use single variable for matched transistors, 
and Table~\ref{tab:5t_corr} summarizes the correlation coefficients of all the parameters, highlighting model accuracy. We show the results of applying the transformer model for three sets of unseen target specifications in Table~\ref{tab:5t_specs}: the optimized circuit can be seen to meet all requirements.

\noindent
\textbf{CM-OTA}
The CM-OTA topology incorporates a differential input stage, succeeded by three current mirror loads. A total of nine devices require sizing in this configuration. The correlation coefficient between the device parameters predicted by the transformer model and the SPICE-based validation data are shown in Table~\ref{tab:cmota_corr} and display high accuracy. Finally, Table~\ref{tab:cmota_specs} delineates the target specifications for three randomly selected designs from the validation set. As in the case of the 5T-OTA, the output of the transformer yields optimized circuits that meet all performance requirements. 

\begin{table}[t]
    \centering
    % \vspace{-2mm}
    % First table
    \begin{minipage}{\linewidth}
        \centering
        \caption{\centering Correlation coefficient of device parameters between validation data and model outputs for the CM-OTA.}
        \resizebox{1\linewidth}{!}{
            \begin{tabular}{|>{\centering\arraybackslash}p{0.9cm}|>{\centering\arraybackslash}p{2.3cm}|c|c|c|c|}
                \hline
                \multirow{2}{*}{\makecell{\textbf{MOS} \\ \textbf{devices}}} & \multirow{2}{*}{\makecell{\textbf{Transistor} \\ \textbf{information}}} & \multicolumn{4}{c|}{\makecell{\textbf{Correlation coefficient}}} \\ \cline{3-6}
                & & \textbf{$g_m$} & \textbf{$g_{ds}$} & \textbf{$C_{ds}$} & \textbf{$C_{gs}$} \\
                \hline
                M1/M2 & Matched CM load & 0.811 & 0.838 & 0.871 & 0.875 \\ \cline{1-6}
                M3/M4 & DP & 0.798 & 0.683 & 0.878 & 0.883 \\ \cline{1-6} 
                M5 & Tail MOS & 0.817 & 0.867 & 0.601 & 0.760 \\ \cline{1-6} 
                M6/M7 & Matched CM load & 0.893 & 0.803 & 0.881 & 0.895 \\ \cline{1-6} 
                M8/M9 & Matched CM load & 0.912 & 0.914 & 0.891 & 0.892 \\ \cline{1-6} 
                \hline
            \end{tabular}
        }
        \label{tab:cmota_corr}
    \end{minipage}
    
    \vspace{0.5mm} % Space between tables

    % Second table
    \begin{minipage}{\linewidth}
        \centering
        \caption{\centering Comparison of optimized design performance with target specifications for the CM-OTA}
        \resizebox{1\linewidth}{!}{
            \begin{tabular}{|c|c|c|c|c|c|}
                \hline
                \multicolumn{2}{|c|}{\textbf{Gain (dB)}} & \multicolumn{2}{c|}{\makecell{\textbf{UGF (MHz)}}} & \multicolumn{2}{c|}{\makecell{\textbf{3dB bandwidth (MHz)}}} \\ \cline{1-6}
                \makecell{\textbf{Target}} & \makecell{\textbf{Optimized}} & \makecell{\textbf{Target}} & \makecell{\textbf{Optimized}} & \makecell{\textbf{Target}} & \makecell{\textbf{Optimized}} \\
                \hline
                20.83 & 21.99 & 345.9 & 475.74 & 30.84 & 37.65 \\ \cline{1-6} 
                21.55 & 23.25 & 247.98 & 408.11 & 20.15 & 27.48 \\ \cline{1-6} 
                23.8 & 24.3 & 1033.77 & 1478.5 & 71.47 & 104.24 \\ \cline{1-6} 
                \hline
            \end{tabular}
        }
        \label{tab:cmota_specs}
    \end{minipage}

    \vspace{-4mm}
\end{table}

\noindent
\textbf{2S-OTA}
The 2S-OTA topology includes a 5T-OTA in the first stage, followed by a common source amplifier comprising seven devices. Table~\ref{tab:2sota_corr} provides a summary of the correlation coefficient between the device parameters predicted by the transformer model and those generated by SPICE, thereby affirming the accuracy of the model. Furthermore, Table~\ref{tab:2sota_specs} presents the target specifications for three randomly selected designs from the validation set. Again, the transformer delivers optimized circuits that meet all specifications.

\begin{table}[b]
    \centering
    \vspace{-4mm}
    % First table
    \hspace*{-0.03\linewidth}
    \begin{minipage}{1.0\linewidth}
        \centering
        \caption{\centering Correlation coefficient of device parameters between validation data and model outputs for the 2S-OTA.}
        \resizebox{1\linewidth}{!}{
            \begin{tabular}{|>{\centering\arraybackslash}p{0.9cm}|>{\centering\arraybackslash}p{2.3cm}|c|c|c|c|}
                \hline
                \multirow{2}{*}{\makecell{\textbf{MOS} \\ \textbf{devices}}} & \multirow{2}{*}{\makecell{\textbf{Transistor} \\ \textbf{information}}} & \multicolumn{4}{c|}{\makecell{\textbf{Correlation coefficient}}} \\ \cline{3-6}
                & & \textbf{$g_m$} & \textbf{$g_{ds}$} & \textbf{$C_{ds}$} & \textbf{$C_{gs}$} \\
                \hline
                M1/M2 & 1\textsuperscript{st} stage active load & 0.942 & 0.936 & 0.876 & 0.879 \\ \cline{1-6}
                M3/M4 & 1\textsuperscript{st} stage DP & 0.988 & 0.945 & 0.913 & 0.915 \\ \cline{1-6} 
                M5 & 1\textsuperscript{st} stage tail MOS & 0.928 & 0.989 & 0.918 & 0.922 \\ \cline{1-6} 
                M6 & 2\textsuperscript{nd} stage tail MOS & 0.856 & 0.881 & 0.843 & 0.798 \\ \cline{1-6} 
                M7 & 2\textsuperscript{nd} stage CS & 0.892 & 0.887 & 0.785 & 0.880 \\ 
                \hline
            \end{tabular}
        }
        \label{tab:2sota_corr}
    \end{minipage}
    
    % \vspace{0.5mm} % Space between tables

    % Second table
    \begin{minipage}{\linewidth}
        \centering
        \caption{\centering Comparison of optimized design performance with target specifications for the 2S-OTA}
        \resizebox{1\linewidth}{!}{
            \begin{tabular}{|c|c|c|c|c|c|}
                \hline
                 \multicolumn{2}{|c|}{\textbf{Gain (dB)}} & \multicolumn{2}{c|}{\makecell{\textbf{UGF (MHz)}}} & \multicolumn{2}{c|}{\makecell{\textbf{3dB bandwidth (kHz)}}} \\ \cline{1-6}
                \makecell{\textbf{Target}} & \makecell{\textbf{Optimized}} & \makecell{\textbf{Target}} & \makecell{\textbf{Optimized}} & \makecell{\textbf{Target}} & \makecell{\textbf{Optimized}} \\
                \hline
                43.6 & 45.61 & 13.33 & 13.4 & 90 & 140 \\ \cline{1-6} 
                47.17 & 47.93 & 11.09 & 11.77 & 80 & 90 \\ \cline{1-6} 
                55.19 & 46.04 & 9.42 & 10.11 & 60 & 91 \\ 
                \hline
            \end{tabular}
        }
        \label{tab:2sota_specs}
    \end{minipage}

    \vspace{-2mm}
\end{table}

From the correlation coefficient analysis, we observe that in some cases the coefficients can be relatively lower (e.g., $<$0.8). These cases correspond to scenarios where the corresponding parameter does not impact the performance metrics significantly, while the other parameter has a more dominant influence. This behavior is attributed to the attention mechanism of the transformer which weights the importance of different parameters based on their level of influence on the performance specifications. As a result, the contributions of less impactful parameters may be overshadowed, leading to a lower correlation coefficient in those specific cases. 

% \bluefn{When I read this entire results section, all I see in the text is ``5T-OTA'': you don't even mention the other OTAs. If the reviewer is in a hurry, s/he will think you really have only done 5T-OTAs. Need to write this better to bring out the other two OTA types. Your paper is already weak because of the simplicity of the circuits (compare with any of the other papers on OTA sizing, which show much more complex circuits). Don't shoot yourself in the foot even further.}

\noindent
%\redHL{~{\em 2) LUT-based transistor width estimation.} For each OTA topology,\bluefn{List them by name so that the reader does not just see ``5T OTA'' in the results section.} we use our approach to determine the transistor sizes for 100 distinct performance specifications that are unseen in the training set. For each set of performance specifications, the transformer model predicts the circuit parameters, which are translated to transistor sizes using the LUT-based method. We report the performance of the design based on a Spectre simulation of the sized OTA circuit. Table~\ref{tab:specs} shows the list of target specs and obtained specs\bluefn{No! You don't \underline{obtain} a spec. See earlier comment. \textbf{FIXED}} for three designs from the validation set for each topology. \textbf{IN MY OPINION, WE DON'T NEED THIS RED PART ANYMORE}}

\begin{table}[t]
    % \vspace{-0.2cm}
    \caption{Runtime analysis of training and inferencing stages.} 
    % \blueHL{This table is poorly explained. You never say anywhere that you run 100 designs. I have changed 95 to 95/100, etc., to make this more obvious in the paper, but you also need to say this in the text.  Please read the paper adversarially, from the point of view of the reviewer. Right now, there are many items that will provoke instant rejection from the reviewer, and you have not bothered to try and address these. I've caught what I could, but I am sure there are other issues because I am starting from almost zero. \textbf{NOTED}}}
    \centering
    \resizebox{1\linewidth}{!}{\begin{tabular}{|c|c|c|c|c|c|c|c|}
        \hline
        \multirow{2}{*}{\makecell{\textbf{OTA} \\ \textbf{topology}}} & \multirow{2}{*}{\makecell{\textbf{One-time} \\\textbf{training} \\\textbf{duration}}}  & \multicolumn{2}{l|}{\makecell{\textbf{Single iteration}}} & \multicolumn{3}{l|}{\makecell{\textbf{Multiple iterations}}} \\ \cline{3-7}
        & & \makecell{\textbf{\#designs}\\ \textbf{optimized}} & \makecell{\textbf{Average} \\\textbf{time}} & \makecell{\textbf{\#designs} \\ \textbf{optimized}} &  \makecell{\textbf{Average} \\\textbf{time}} & \makecell{\textbf{Average}\\\textbf{\#iterations}} \\
        \hline
        5T-OTA & 8.5h  & 95/100 & 37s & 5/100 & 111s & 3\\
        \hline
        CM-OTA & 22h  & 98/100 & 46s & 2/100 & 230s & 5\\
        \hline
        2S-OTA & 11h & 90/100 & 36s & 10/100 & 180s & 5\\
        \hline
    \end{tabular}}
    \label{tab:runtime}
    \vspace{-5mm}
\end{table}

% \vspace{-2mm}

\subsection{Runtime analysis}
\noindent
Table~\ref{tab:runtime} provides a detailed runtime analysis, including both the one-time SPICE-based training duration and the average runtime per design optimization by the trained model. The reported runtime encompasses the entire process, from sequence inference by the trained transformer model (taking approximately 0.5s per sequence)
% -- where each sequence takes approximately 0.5 seconds -- 
to the LUT-based estimation and subsequent SPICE simulation verification. 
In cases where performance criteria are not fully met due to minor prediction inaccuracies, a ``copilot'' mode is activated, performing iterative refinements with progressively tighter specifications to introduce a design margin that compensates for errors. This ensures that all design specifications are ultimately satisfied, typically requiring only a few additional iterations, thereby balancing model accuracy with computational efficiency to achieve reliable design convergence. The runtime ranges from just above 30s to just under four minutes, significantly lower than competing methods.
% \vspace{-2mm}

\subsection{Qualitative comparison with prior approaches.}

\noindent
Table~\ref{tab:comparison} compares our approach for OTA sizing with prior methods, including simulated annealing (SA)~\cite{gielen_90}, particle swarm optimization (PSO)~\cite{vural_12}, graph convolutional network-based RL (GCN-RL)~\cite{Wang_2020}, weighted expected improvement-based Bayesian optimization (WEIBO)~\cite{lyu_18}, and differential evolutionary (DE) algorithm~\cite{liu_09}.

We utilize various metrics for comparison. \textit{SPICE simulation dependency} gauges the reliance on costly simulations for convergence: lower dependence indicates greater efficiency. \textit{Sizing accuracy} measures how well the approach satisfies all design specifications.
% \textit{Optimization efficiency} captures the trade-off between resource usage and the ability to find optimal designs, while sizing accuracy evaluates how reliably each method determines correct transistor sizes.
% \redfn{What is your basis for saying this method is high? You never evaluate whether the sizing is performed at low area or power -- so what is your metric for deciding that this is true? \textbf{I probably thought about this metric in the wrong way. I thought of it as a function of the performance of the optimized design and computational cost for the optimization.} {\em So how did you think about it? What is your basis for saying that the optimization efficiency is high? Can you use a different term that captures what you were trying to say?} \textbf{I feel like this metric is kind of redundant now. We have SPICE simulation dependency metric and Convergence cost, which pretty much covers this metric as well.} {\em I am ok with removing this row of the table.} RESOLVED (from my end).} 
\textit{Runtime} reflects the time required to reach a solution, and \textit{memory utilization} pertains to the amount of memory resources consumed during the optimization process.
Our approach, using a trained transformer model with precomputed LUTs, significantly reduces SPICE simulation dependency -- achieving over 90\% of sizing without simulations -- while improving accuracy and reducing runtime from hours to seconds, positioning it as a highly efficient solution.

\begin{table}[ht]
    \centering
    \vspace{-2mm}
    \caption{Qualitative comparison with prior sizing approaches. }
    % \redHL{Why are some items bf and others not? If we are trying to bf our approach, why is sizing accuracy = high not in bf?} \textbf{Because, its high for others also, so nothing special in our method.}}
    \resizebox{1\linewidth}{!}{
    \begin{tabular}{|c|c|c|c|c|c|c|}
        \hline
        \makecell{\textbf{Sizing method}} & \makecell{\cite{gielen_90}} & 
        \makecell{\cite{vural_12}} &
        \makecell{\cite{Wang_2020}} &
        \makecell{\cite{lyu_18}} &  
         \makecell{\cite{liu_09}} &
         \makecell{\textbf{Our approach}}
         \\
        \hline
        \textbf{Algorithm} & SA & PSO & \makecell{GCN \\ + RL} & WEIBO & DE & \makecell{Transformer \\ + LUT } \\
        \hline
        \makecell{\textbf{SPICE simulation} \\ \textbf{dependency}} & \makecell{Very high} & \makecell{Very high} & \makecell{Low to \\ moderate}& \makecell{High} &  \makecell{Very high} & \makecell{\textbf{Very low\textsuperscript{*}}}\\
        % \hline
        % \makecell{\textbf{Optimization} \\ \textbf{efficiency}} & \makecell{Moderate \\ to low} & \makecell{Moderate \\ to low} & Moderate &  High & High & High\\
        \hline
        \makecell{\textbf{Sizing accuracy}}& Variable & \makecell{Moderate \\ to high}& High & High & High & \makecell{ High}\\
        \hline
        \makecell{\textbf{Runtime}} & \makecell{Very slow}& \makecell{Very slow}& \makecell{Moderate}& \makecell{Moderate}  & \makecell{Slow}& \makecell{\textbf{Very fast}} \\
        \hline
        \makecell{ \textbf{Memory} \\ \textbf{utilization}} & Moderate & Moderate & \makecell{Moderate \\ to high}& \makecell{High} & \makecell{Very high} & \makecell{\textbf{Moderate} \\ \textbf{to low}}\\

        \hline
    \end{tabular}}
    \label{tab:comparison}
    \scriptsize\textsuperscript{*} $>$90\% of sizing is performed without SPICE simulations, as shown in Table~\ref{tab:runtime}.
    \vspace{-3mm}
\end{table}

\section{Conclusion}
\label{sec:conclusion}

\noindent
We have introduced a fully automated rapid-sizing tool for OTA circuits that utilizes a transformer-based attention mechanism. Our framework successfully meets stringent design specifications across multiple unseen designs for three distinct OTA topologies, achieving a success rate exceeding 90\% of the designs on the first attempt. Unlike much prior research, our method eliminates the need for extensive and costly SPICE simulations for each design, enhancing computational efficiency and accelerating the design process for OTA circuits. 

\newpage
% \bstctlcite{IEEEexample:BSTcontrol}
% \bibliographystyle{alpha}
% \bibliographystyle{misc/ieeetr2}
\bibliographystyle{misc/IEEEtran}

\begin{thebibliography}{10}
\providecommand{\url}[1]{#1}
\csname url@samestyle\endcsname
\providecommand{\newblock}{\relax}
\providecommand{\bibinfo}[2]{#2}
\providecommand{\BIBentrySTDinterwordspacing}{\spaceskip=0pt\relax}
\providecommand{\BIBentryALTinterwordstretchfactor}{4}
\providecommand{\BIBentryALTinterwordspacing}{\spaceskip=\fontdimen2\font plus
\BIBentryALTinterwordstretchfactor\fontdimen3\font minus \fontdimen4\font\relax}
\providecommand{\BIBforeignlanguage}[2]{{%
\expandafter\ifx\csname l@#1\endcsname\relax
\typeout{** WARNING: IEEEtran.bst: No hyphenation pattern has been}%
\typeout{** loaded for the language `#1'. Using the pattern for}%
\typeout{** the default language instead.}%
\else
\language=\csname l@#1\endcsname
\fi
#2}}
\providecommand{\BIBdecl}{\relax}
\BIBdecl

\bibitem{harjani_89}
R.~Harjani, R.~Rutenbar, and L.~Carley, ``{OASYS}: A framework for analog circuit synthesis,'' \emph{IEEE Transactions on Computer-Aided Design of Integrated Circuits and Systems}, vol.~8, no.~12, pp. 1247--1266, Dec. 1989.

\bibitem{koza_96}
J.~R. Koza, F.~H. Bennett, D.~Andre, and M.~A. Keane, ``Automated design of both the topology and sizing of analog electrical circuits using genetic programming,'' in \emph{Artificial Intelligence in Design '96}, J.~S. Gero and F.~Sudweeks, Eds.\hskip 1em plus 0.5em minus 0.4em\relax Dordrecht, Netherlands: Springer, 1996, pp. 151--170.

\bibitem{Kruiskamp_95}
W.~Kruiskamp and D.~Leenaerts, ``{DARWIN}: {CMOS} opamp synthesis by means of a genetic algorithm,'' in \emph{Proceedings of the ACM/IEEE Design Automation Conference}, 1995, pp. 433--438.

\bibitem{gielen_90}
G.~Gielen, H.~Walscharts, and W.~Sansen, ``Analog circuit design optimization based on symbolic simulation and simulated annealing,'' \emph{IEEE Journal of Solid-State Circuits}, vol.~25, no.~3, pp. 707--713, Jun. 1990.

\bibitem{vural_12}
R.~A.~Vural and T.~Yildirim, ``Analog circuit sizing via swarm intelligence,'' \emph{AEU -- International Journal of Electronics and Communications}, vol.~66, p. 732–740, Sep. 2012.

\bibitem{abel_22}
I.~Abel and H.~Graeb, ``{FUBOCO}: Structure synthesis of basic op-amps by functional block composition,'' \emph{ACM Transactions on Design Automation of Electronic Systems}, vol.~27, no.~6, Jun. 2022.

\bibitem{abel_22_2}
I.~Abel, M.~Neuner, and H.~E. Graeb, ``A hierarchical performance equation library for basic op-amp design,'' \emph{IEEE Transactions on Computer-Aided Design of Integrated Circuits and Systems}, vol.~41, no.~7, pp. 1976--1989, 2022.

\bibitem{hershenson_01}
M.~Hershenson, S.~Boyd, and T.~Lee, ``Optimal design of a {CMOS} op-amp via geometric programming,'' \emph{IEEE Transactions on Computer-Aided Design of Integrated Circuits and Systems}, vol.~20, no.~1, pp. 1--21, Jan. 2001.

\bibitem{budak_21}
A.~F. Budak, P.~Bhansali, B.~Liu, N.~Sun, D.~Z. Pan, and C.~V. Kashyap, ``{DNN-Opt}: An {RL} inspired optimization for analog circuit sizing using deep neural networks,'' in \emph{Proceedings of the ACM/IEEE Design Automation Conference}, 2021, pp. 1219--1224.

\bibitem{settaluri_20}
K.~Settaluri, A.~Haj-Ali, Q.~Huang, K.~Hakhamaneshi, and B.~Nikolic, ``{AutoCkt}: Deep reinforcement learning of analog circuit designs,'' in \emph{Proceedings of the Design, Automation \& Test in Europe}, 2020, pp. 490--495.

\bibitem{Wang_2020}
H.~Wang, K.~Wang, J.~Yang, L.~Shen, N.~Sun, H.-S. Lee, and S.~Han, ``{GCN-RL} circuit designer: Transferable transistor sizing with graph neural networks and reinforcement learning,'' in \emph{Proceedings of the ACM/IEEE Design Automation Conference}, 2020, pp. 1--6.

\bibitem{choi_23}
M.~Choi, Y.~Choi, K.~Lee, and S.~Kang, ``Reinforcement learning-based analog circuit optimizer using {$g_m/I_D$} for sizing,'' in \emph{Proceedings of the ACM/IEEE Design Automation Conference}, 2023.

\bibitem{vaswani_17}
A.~Vaswani, N.~Shazeer, N.~Parmar, J.~Uszkoreit, L.~Jones, A.~N. Gomez, L.~Kaiser, and I.~Polosukhin, ``Attention is all you need,'' in \emph{Advances in Neural Information Processing Systems}, vol.~30, Dec. 2017, pp. 5998--6008.

\bibitem{ochoa_98}
A.~Ochoa, ``A systematic approach to the analysis of general and feedback circuits and systems using signal flow graphs and driving-point impedance,'' \emph{IEEE Transactions on Circuits and Systems II}, vol.~45, no.~2, pp. 187--195, Feb. 1998.

\bibitem{schmid_18}
H.~Schmid and A.~Huber, ``Analysis of switched-capacitor circuits using driving-point signal-flow graphs,'' \emph{Analog Integrated Circuits and Signal Processing}, vol.~96, pp. 495--507, Sep. 2018.

\bibitem{Mason53}
S.~J. Mason, ``Feedback theory-some properties of signal flow graphs,'' \emph{Proceedings of the IRE}, vol.~41, no.~9, pp. 1144--1156, 1953.

\bibitem{schmid_yt}
{H. Schmid}, ``{{HT FHNW EIT}: Analog and mixed-signal circuits and signal processing},'' \url{https://tube.switch.ch/channels/d206c96c}.

\bibitem{rico_16}
R.~Sennrich, B.~Haddow, and A.~Birch, ``Neural machine translation of rare words with subword units,'' in \emph{Annual Meeting of the Association for Computational Linguistics}, Aug. 2016, pp. 1715--1725.

\bibitem{silviera_96}
F.~Silveira, D.~Flandre, and P.~Jespers, ``A {$g_m$/$I_D$} based methodology for the design of {CMOS} analog circuits and its application to the synthesis of a silicon-on-insulator micropower {OTA},'' \emph{IEEE Journal of Solid-State Circuits}, vol.~31, no.~9, pp. 1314 -- 1319, Oct. 1996.

\bibitem{jespers_17}
P.~Jespers and B.~Murmann, \emph{Systematic Design of Analog {CMOS} Circuits: Using Pre-Computed Lookup Tables}.\hskip 1em plus 0.5em minus 0.4em\relax Cambridge, UK: Cambridge University Press, 2017.

\bibitem{lyu_18}
W.~Lyu, P.~Xue, F.~Yang, C.~Yan, Z.~Hong, X.~Zeng, and D.~Zhou, ``An efficient {Bayesian} optimization approach for automated optimization of analog circuits,'' \emph{IEEE Transactions on Circuits and Systems I}, vol.~65, no.~6, pp. 1954--1967, Jun. 2018.

\bibitem{liu_09}
B.~Liu, Y.~Wang, Z.~Yu, L.~Liu, M.~Li, Z.~Wang, J.~Lu, and F.~V. Fernández, ``Analog circuit optimization system based on hybrid evolutionary algorithms,'' \emph{Integration}, vol.~42, no.~2, pp. 137--148, Apr 2009.

\end{thebibliography}
\end{document}